\documentclass{PoS}

\title{Disconnected quark loop contributions to nucleon observables using $N_f=2$ twisted clover fermions at the physical value of the light quark mass}

\ShortTitle{Disconnected contributions at the physical quark mass}

\author{Abdou Abdel-Rehim$^a$, Constantia Alexandrou$^{ab}$,Martha Constantinou$^{ab}$, Kyriakos Hadjiyiannakou$^{ab}$, Karl Jansen$^c$, Christos Kallidonis$^a$, Giannis Koutsou$^a$, \speaker{Alejandro Vaquero Avil\'es-Casco}$^d$\\
     \\
     \llap{$^a$}Computation-based Science and Technology Research Center, CaSToRC\\
     The Cyprus Institute, 20 Kavafi Str. Nicosia 2121, Cyprus\\
     \\
     \llap{$^b$}Department of Physics, University of Cyprus\\
     P.O. Box 20537, 1678 Nicosia, Cyprus\\
     \\
     \llap{$^c$}NIC, DESY\\
     Platanenallee 6, D-15738 Zeuthen, Germany\\
     \\
     \llap{$^d$}INFN Sezione di Milano-Bicocca\\
     Edificio U2, Piazza della Scienza 3, 20126 Milano, Italy\\
     \\
     E-mail: \email{a.abdel-Rehim@cyi.ac.cy}, \email{alexand@ucy.ac.cy}, \email{marthac@ucy.ac.cy}, \email{hadjigiannakou.kyriakos@ucy.ac.cy}, \email{karl.jansen@desy.de}, \email{c.kallidonis@cyi.ac.cy}, \email{g.koutsou@cyi.ac.cy}, \email{Alejandro.Vaquero@mib.infn.it}}

\abstract{We compute the disconnected quark loops contributions entering the determination of  nucleon observables, by using a $N_f = 2$ ensemble of twisted mass fermions with a
	  clover term at a pion mass $m_\pi = 133$ MeV. We employ exact deflation and implement all calculations in GPUs, enabling us to  achieve large statistics and a good signal.}

\FullConference{The 33rd International Symposium on Lattice Field Theory\\
		14 -18 July 2015\\
		Kobe International Conference Center, Kobe, Japan}

\begin{document}

\section{Introduction}
The evaluation of disconnected diagrams has historically been extremely difficult in lattice QCD computations. The main obstacle is the calculation of the all-to-all propagator, which is unfeasible with current computer power.
However, the situation in the last years has improved dramatically. Although their computation is still a difficult task, improved methods and new computer architectures have made it attainable \cite{yoRes, hProb, bali}. In this
work we examine the disconnected diagrams calculated directly on a $N_f=2$ ensemble of twisted-clover fermions at the physical value of the light quark mass. To this end we benefit from several algorithmic improvements,
as well as for the increasing computer power available to us.

\section{Methods for disconnected diagrams}
\label{sec:disM}
\subsection{Stochastic estimators\label{subs:stoc}}
A direct computation of the inverse $G$ of the fermionic matrix $M$ is prohibitely expensive, hence we must rely on approximations. The main tool for the all-to-all computations is the stochastic estimator
\cite{stoch}, which allows to compute an unbiased estimation of the propagator by generating a set of $N$ white-noise sources $\left|\eta_j\right\rangle$. Each entry of the source is filled with random noise, $Z_4$
noise in our case. Thus generated sources comply with the following properties,

\begin{equation} 
\frac{1}{N}\sum_{j=1}^{N}\left|\eta_j\right\rangle = O\left(\frac{1}{\sqrt{N}}\right),\quad\frac{1}{N}\sum^{N}_{j=1}\left|\eta_j\right\rangle\left\langle\eta_j\right| = \mathbb{I} + O\left(\frac{1}{\sqrt{N}}\right). \label{eqN}
\end{equation} 
The first property ensures that our estimate of the propagator is unbiased. The second one
allows us to reconstruct the inverse matrix by solving for $\left|s_j\right\rangle$ in
\begin{eqnarray}
M\left|s_j\right\rangle = \left|\eta_j\right\rangle\quad\longrightarrow
\label{etaToS}
&\quad G_E:=\frac{1}{N}\sum_{j=1}^{N}\left|s_j\right\rangle
\left\langle\eta_j\right|\approx G.
\label{estiM}
\end{eqnarray}
The deviation of our estimator from the exact solution is given by
\begin{equation}
G-G_E = G\times\left(\mathbb{I}-
\frac{1}{N}\sum_{j=1}^{N}\left|\eta_j\right\rangle\left\langle\eta_j\right|\right) \approx O\left(\frac{1}{\sqrt{N}}\right),
\label{errS}
\end{equation}
which has only off-diagonal entries, a nice property ensured by the usage of $Z_N$ noise sources \cite{ZN}.

\subsection{The Truncated Solver Method (TSM)}

Clearly, stochastic estimators are not enough. Each inversion is expensive, while the error decreases as $1/\sqrt{N}$. The Truncated Solver Method (TSM)~\cite{TSM} enables us to increase $N$ at a reduced cost: instead
of solving to high precision (HP) Eq.~(\ref{estiM}), we can obtain a low precision (LP) estimate where the inverter, a CG solver in this work, is truncated. The truncation criterion can be a large value of the residual
$\rho_{LP} >> \rho_{HP}$, or a fixed number of iterations $n_{LP} << n_{HP}$. Generation of new sources becomes cheap, however we are introducing a bias by not allowing the inverter to converge. The bias is stochastically
corrected by generating another set of stochastic sources, inverting them to both high and low precision, and computing the difference in the estimates

\begin{equation}
G_{E_{TSM}}:=\underbrace{\frac{1}{N_{\rm HP}}\sum_{j=1}^{N_{\rm HP}}
\left[\left|s_j\right\rangle_{HP} - \left|s_j\right\rangle_{LP}\right]
\left\langle\eta_j\right|}_{\textrm{Correction}} + \underbrace{\frac{1}{N_{\rm LP}}
\sum_{j=N_{\rm HP}+1}^{N_{\rm HP}+N_{\rm LP}}\left|s_j\right\rangle_{LP}
\left\langle\eta_j\right|}_{\textrm{Biased estimate}},
\label{estiTSM}
\end{equation}
which requires $N_{\rm HP}$ high precision inversions and $N_{\rm HP}+N_{\rm LP}$ low precision inversions. If enough sources are used for the correction, the error of this improved estimator scales as
$\propto 1/\sqrt{N_{\rm LP}}$.

The TSM works as long as the LP sources are highly correlated with the HP ones. In this case, the correction is some value that slowly changes as we add new stochastic sources, therefore only a few HP sources are needed to
compute it to high accuracy. This fact offloads the main computational cost onto the LP inversions, which are cheaper. Of course, some tuning is required to ensure that no bias remain after the inversions are truncated.

\begin{figure}[h!]
   \begin{center}
      \begin{minipage}{0.57\linewidth}
        \includegraphics[width=\linewidth,angle=0]{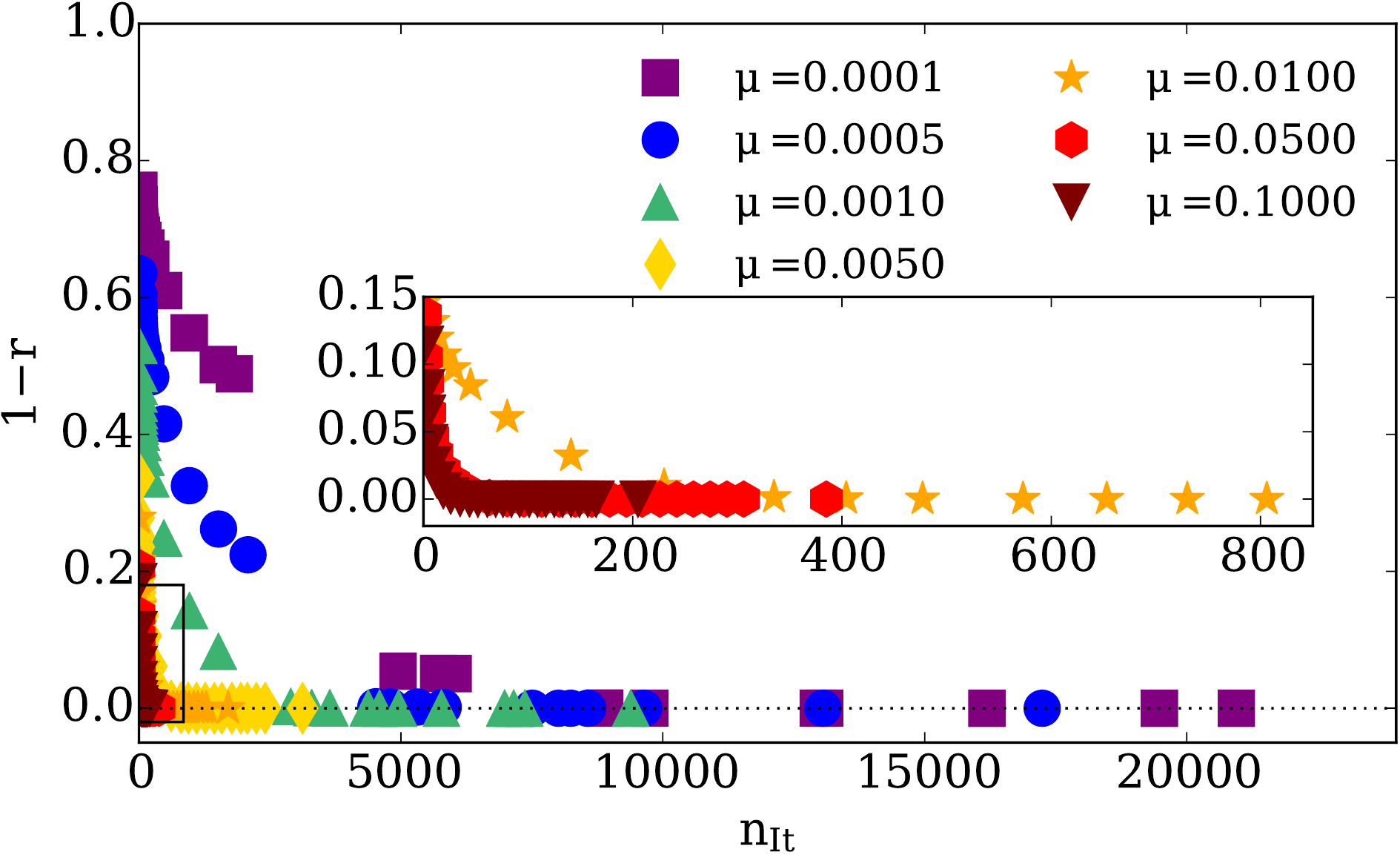} 
      \end{minipage}
      \hspace{0.05\linewidth}
      \begin{minipage}{0.33\linewidth}
        \includegraphics[width=\linewidth,angle=0]{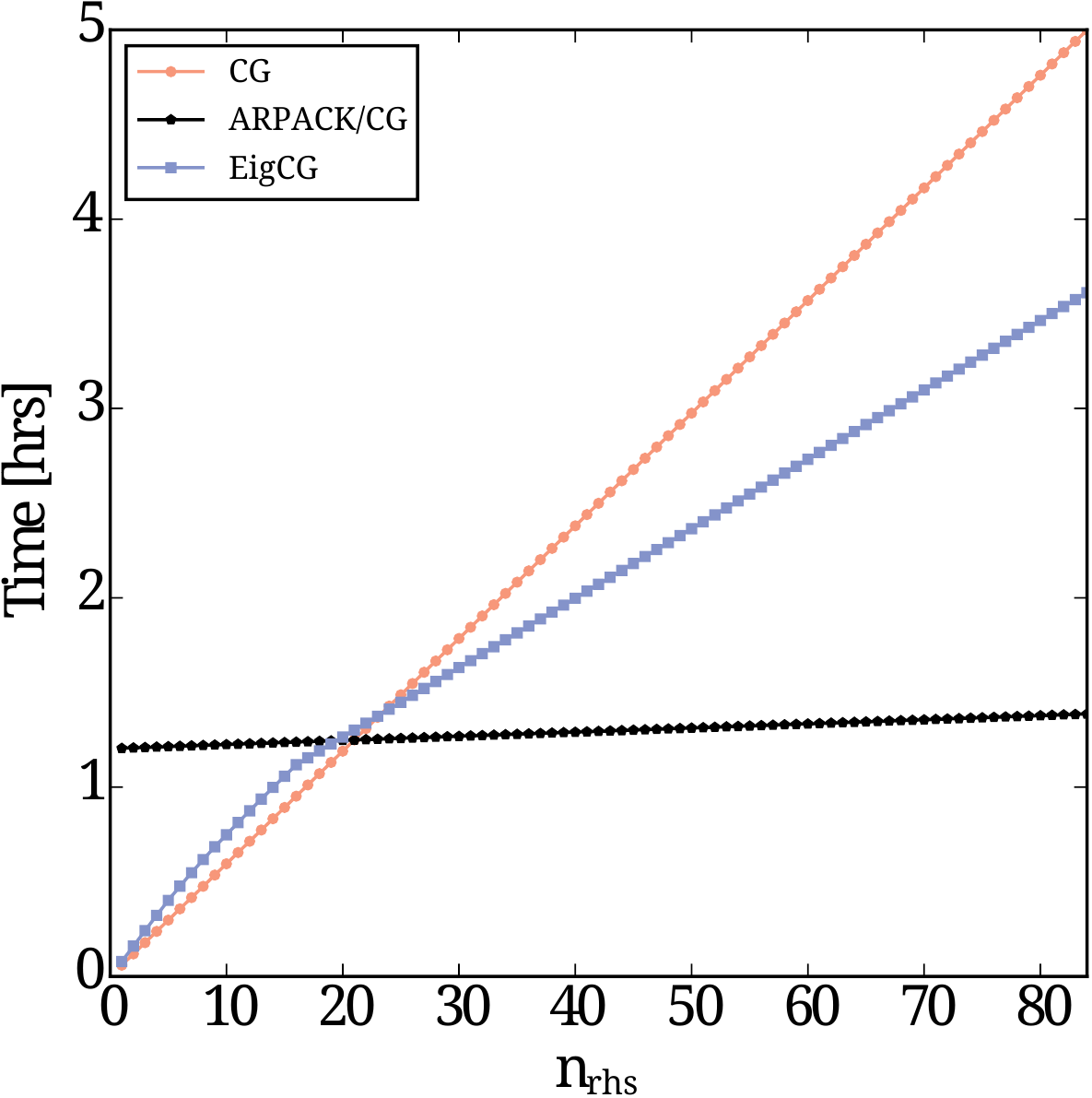}
      \end{minipage}
      \caption{\footnotesize Left: Correlation-cost plot between HP and LP sources in a $24^3\times 48$ twisted-mass configuration for several quark masses $\mu$, where $1-r=0$ means $100\%$ correlation. To guide the reader,
      $m_{u,d}\leq 0.0005$, $m_s\sim 0.01$ and $m_c\sim 0.1$. Right: Deflation performance as the number of eigenvectors increases, compared to eigCG or standard CG.\label{corrDef}}
   \end{center}
\end{figure}

The approach works best for light quark masses, but these must not be very light. As noted in earlier works \cite{yoCp}, if the mass is too large, the inverter is already extremely fast, and most of the computational cost comes
from other sources (contractions, source generation, etc), for which the TSM offers no solution. On the other hand, for too light masses performance drops, due to the fact that the cost to preserve the correlation between HP and
LP increases extremely fast. In the left panel of fig. \ref{corrDef} the cost (in terms of iteration count) is plotted against the correlation $r$ among HP and LP. Although performance gains are still visible, these become
much less noticeable as the mass becomes lighter. It seems to be a sweet spot around the strange quarks mass, but this depends on the overhead of each particular setup.

In this work we employed the TSM as our main variance reduction technique only for the strange disconnected loops. Light loops were computed with exact deflation.

\subsection{Exact deflation}

The TSM fails for light quark masses close to the physical value due to the existence of low-modes. These modes increase the condition number of the fermionic matrix and spoil the convergence of the solver, hence they must be
removed. The low-modes of the twisted-clover operator can be efficiently calculated with an Arnoldi eigensolver, but the fact that $M$ is not a normal operator introduces several complications, i.e., there are different right
and left eigenvectors, and the eigenvectors are not orthogonal, therefore they are calculated with less accuracy than in the normal case. It is much more efficient to work with the hermitian operator $Q=\gamma_5M$ and use a
Lanczos eigensolver. After all, since we use a CG inverter, our operator $M^\dagger M$ shares eigenvectors with $Q$.

Assuming we know the lowest $N_L$ modes of our Dirac operator, we can deflate the inversions to increase the performance of our solver, the higher the value of $N_L$, the faster our inversions become. In principle, we could
make better use of these modes and not only deflate the solver, but also compute exactly the low-mode piece of the propagator

\begin{equation}
Q^{-1} = \sum_i^{N_L}\frac{1}{\lambda_i}\left|v_i\right\rangle\left\langle v_i\right| + \sum_j^N\left|s_j\right\rangle\left\langle\eta_j\right|P_\perp,\quad
P_\perp=\sum_i^{N_L}\left[I-\left|v_i\right\rangle\left\langle v_i\right|\right].
\label{qRecon}
\end{equation}
Unfortunately the usage of even-odd preconditioning in our CG solver does not allow us to proceed this way. The eigenvectors of the even-odd preconditioned twisted-clover operator can't be related to those of the unpreconditioned
one, therefore we need to choose between even-odd preconditioning, or exact reconstruction of the low-modes. The advantages of the exact low-mode reconstruction are obvious, but it also entails
an important disadvantage: the eigenvectors become full vectors, and the storage needs are duplicated, or half the eigenvectors are computed. This implies a double performance decrease of the solver coming from the computation
of less eigenvectors and the removal of the even-odd preconditioning. Since the performance gains of using the full operator were not entirely clear, we stayed with simple deflation plus even-odd preconditioning at
the light quark mass. We are currently investigating the efficiency of the alternative approach.

\subsection{The one-end trick}

The one-end trick was first introduced in two-point meson correlators as a way to reduce noise by implicitly computing a volume sum, thus increasing statistics \cite{vvTrick1,vvTrick2}. The one-end trick works as long as we
have a product of propagators $G^\dagger G$ in our expectation value, which happens in the twisted-mass regularization with $\Gamma$ insertions with an isovector-flavor structure in the twisted basis,

\begin{equation}
\Gamma\left[G_u - G_d\right] = -2i\mu a\Gamma G_d\gamma_5G_u = -2i\mu a\Gamma\gamma_5\sum_{j,k}^N\left|s_j\right\rangle\left\langle 
\eta_j\right|\left.\eta_k\right\rangle\left\langle s_k\right|.
\end{equation}
The one-end trick replaces the inner product $\left\langle \eta_j\right|\left.\eta_k\right\rangle$ by $\delta_{jk}$, performing the volume sum with increased precision and reducing noticeably the errors. The final expresion
is shown in Eq.~\ref{loopVv}, whereas Eq.~\ref{loopStD} shows a similar construction applied to isoscalar operators in the twisted basis.

\hspace{-0.8cm}\begin{minipage}{0.485\linewidth} 
\begin{equation}
\textrm{Tr}\left[\Gamma\left(G_u - G_d\right)\right] = -\frac{2i\mu a}{N}\sum_{j=1}^N \left\langle s_j \Gamma\gamma_5 s_j\right\rangle
\label{loopVv}
\end{equation} 
\end{minipage} 
\begin{minipage}{0.515\linewidth} 
\begin{equation}
\textrm{Tr}\left[\Gamma\left(G_u + G_d\right)\right] = \frac{2}{N}\sum_{j=1}^N \left\langle s_j \gamma_5\Gamma\gamma_5 D_W s\right\rangle 
\label{loopStD}
\end{equation} 
\end{minipage}
Because of the volume sum introduced by our identities, the sources must have entries on all sites, which in turn means that we compute the fermion loop at all insertion times simultaneously.

\section{Ensemble and setup}
\label{sec:setUp}
We calculated the disconnected loops for a $48^3\times 96$ ensemble of $\approx1800$ configurations of $N_f=2$ twisted-clover fermions tuned at the physical value of the light quark mass ($m_\pi\approx 133$ MeV), with
$a = 0.093$ fm. For the light quark loops, we relied on exact deflation to speed-up our CG solver. The implementation of PARPACK has proved very efficient in generating the exact eigenmodes, as shown in fig. \ref{corrDef}, so
we calculated 600 eigenpairs, and our inversion cost became negligible. We used 2250 sources per configuration that were prepared in the CPU, and then transferred to GPUs for efficient inversion, contraction and FFT using the
QUDA library \cite{QUDA,Yo2}. The same eigenvectors were employed for disconnected loop and the connected two-point correlators computation, the latter benefiting greatly from the deflation, since we computed 100 point-to-all
propagators per configuration in randomized positions to increase our statisticsup to 721600 measurements.

For the strange quark disconnected loops, deflation was not employed, but the TSM with 63 HP and 1024 LP sources. In Ref.~\cite{TEA} it was reported that, although deflation can improve performance, at the strange quark
mass the benefits are already fading, due to the inversions becoming fast. This was accentuated in our case with the usage of GPUs. Our statistics in this case were 739600.

\section{Results}
\label{sec:res}
\begin{figure*}[h!]
   \begin{center}
      \begin{minipage}{0.4\linewidth}
        \includegraphics[width=\linewidth,angle=0]{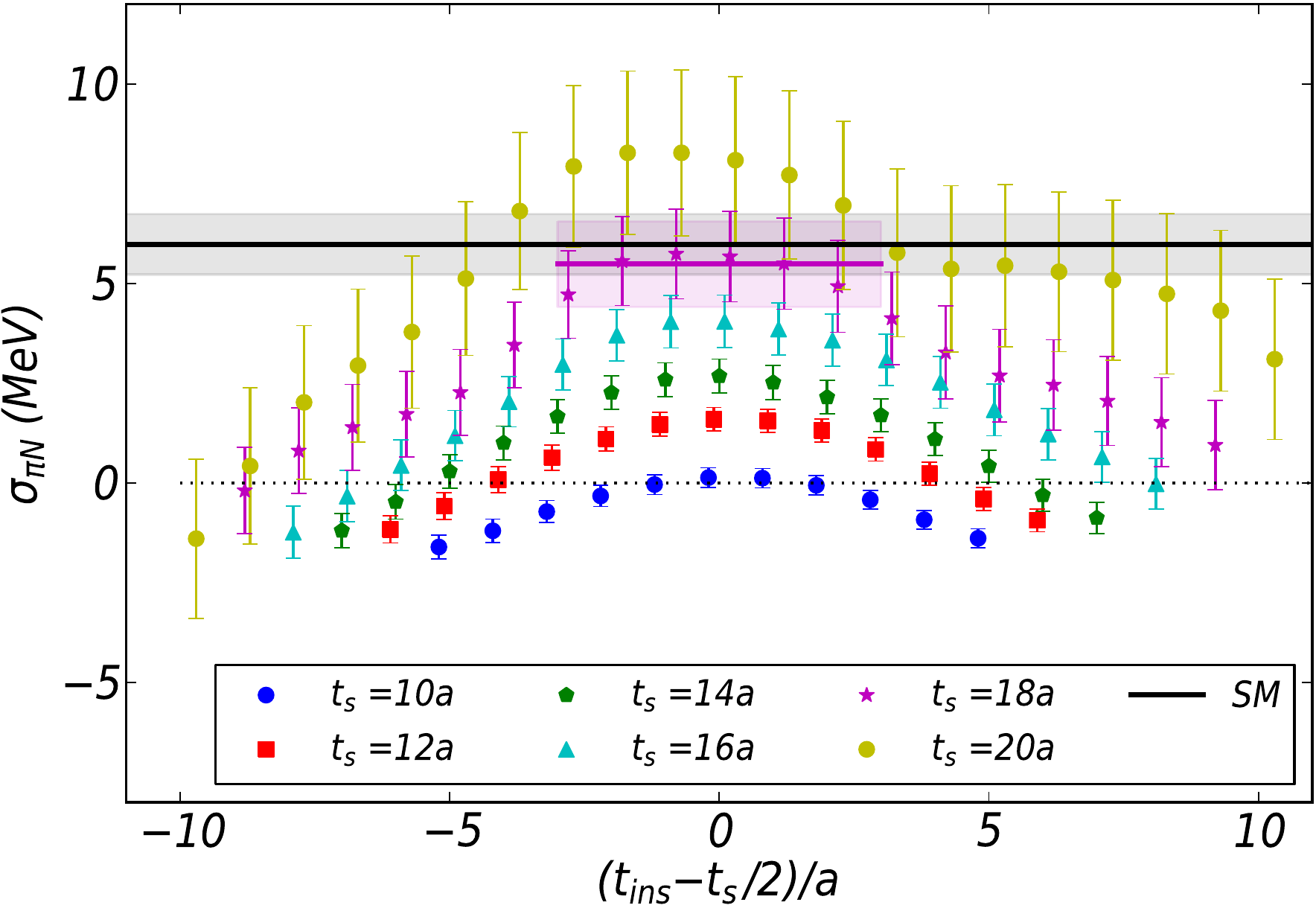}
      \end{minipage}
      \hspace{0.05\linewidth}
      \begin{minipage}{0.4\linewidth}
        \includegraphics[width=\linewidth,angle=0]{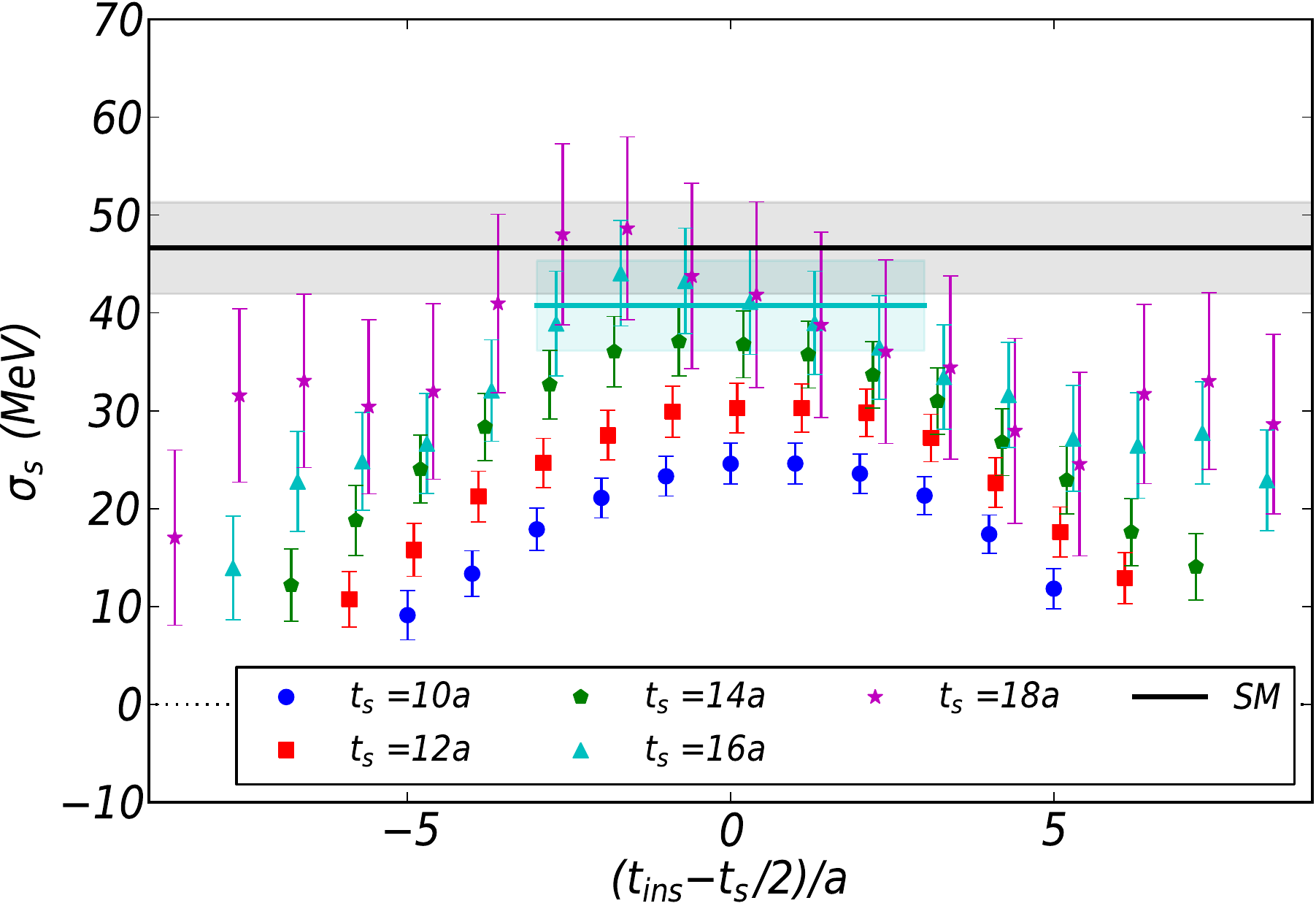}
      \end{minipage}
      \caption{\footnotesize Plateaux for disconnected $\sigma_{\pi N}$ (left) and nucleon $\sigma_s$ (right). The sink time is represented as $t_s$, insertion time as $t_{ins}$, and different colors show different source-sink
      separations. The grey band is the summation method result.\label{platSig}}
   \end{center}
\end{figure*}

\begin{figure*}[h!]
   \begin{center}
      \begin{minipage}{0.4\linewidth}
        \includegraphics[width=\linewidth,angle=0]{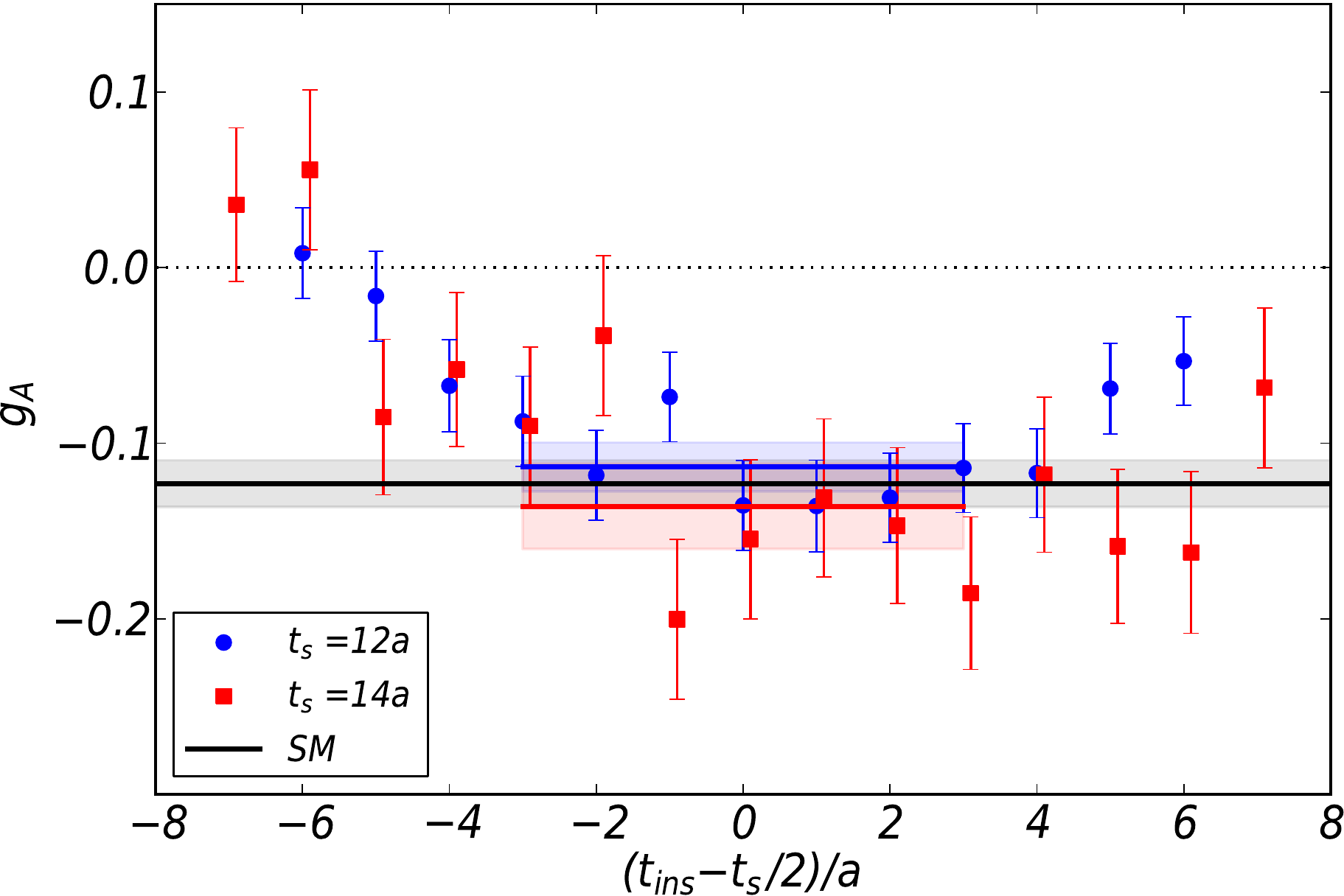} 
      \end{minipage}
      \hspace{0.05\linewidth}
      \begin{minipage}{0.4\linewidth}
        \includegraphics[width=\linewidth,angle=0]{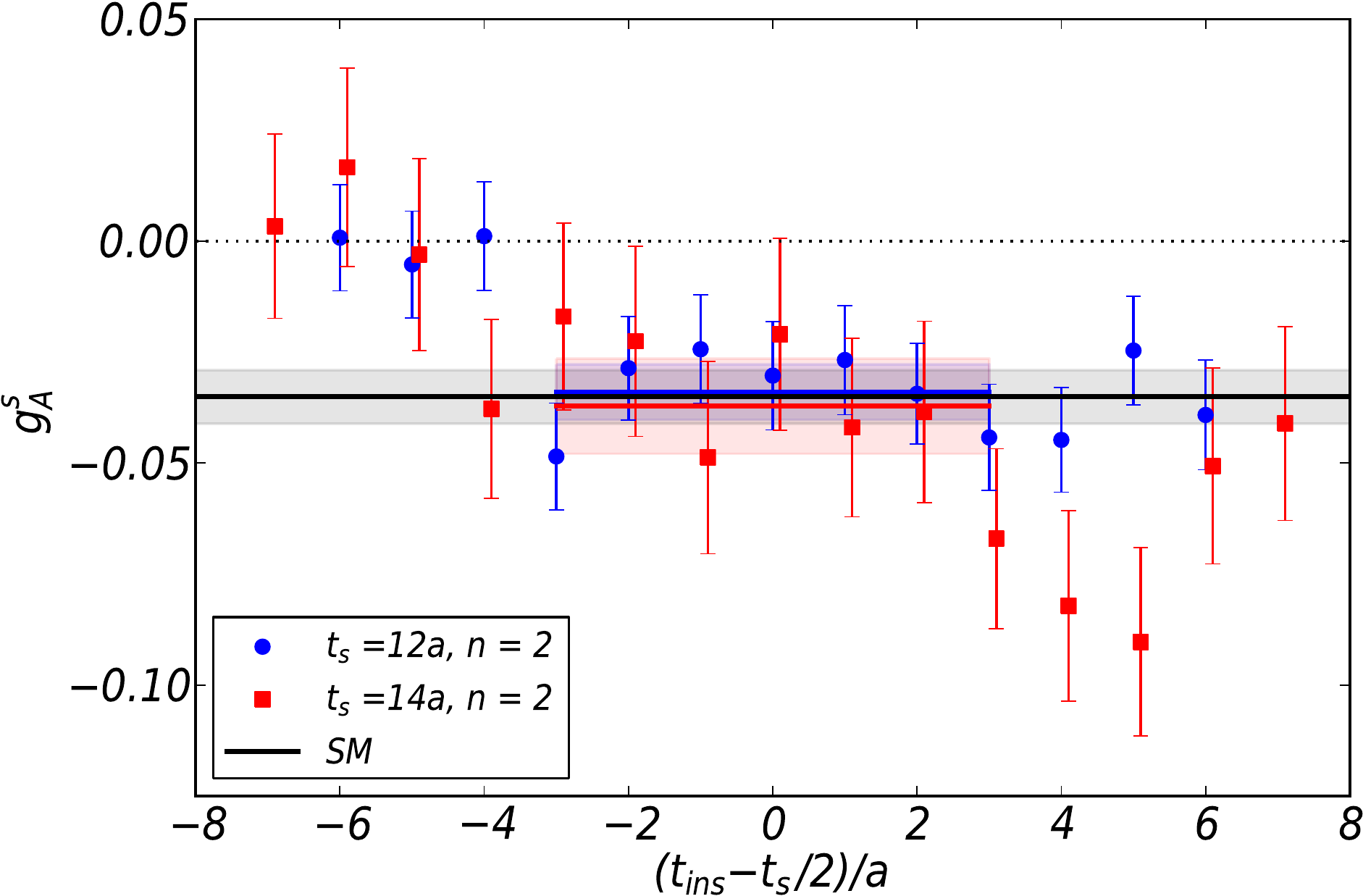} 
      \end{minipage}. 
      \caption{\footnotesize Plateaux for nucleon disconnected $g_A^{u+d}$ (left) and $g_A^s$ (right). The notation is as in the previous plot, and the results are renormalized without considering the $g_A$ mixing between flavors.
      \label{platgA}}
   \end{center}
\end{figure*}

\begin{figure*}[h!]
   \begin{center}
      \begin{minipage}{0.4\linewidth}
        \includegraphics[width=\linewidth,angle=0]{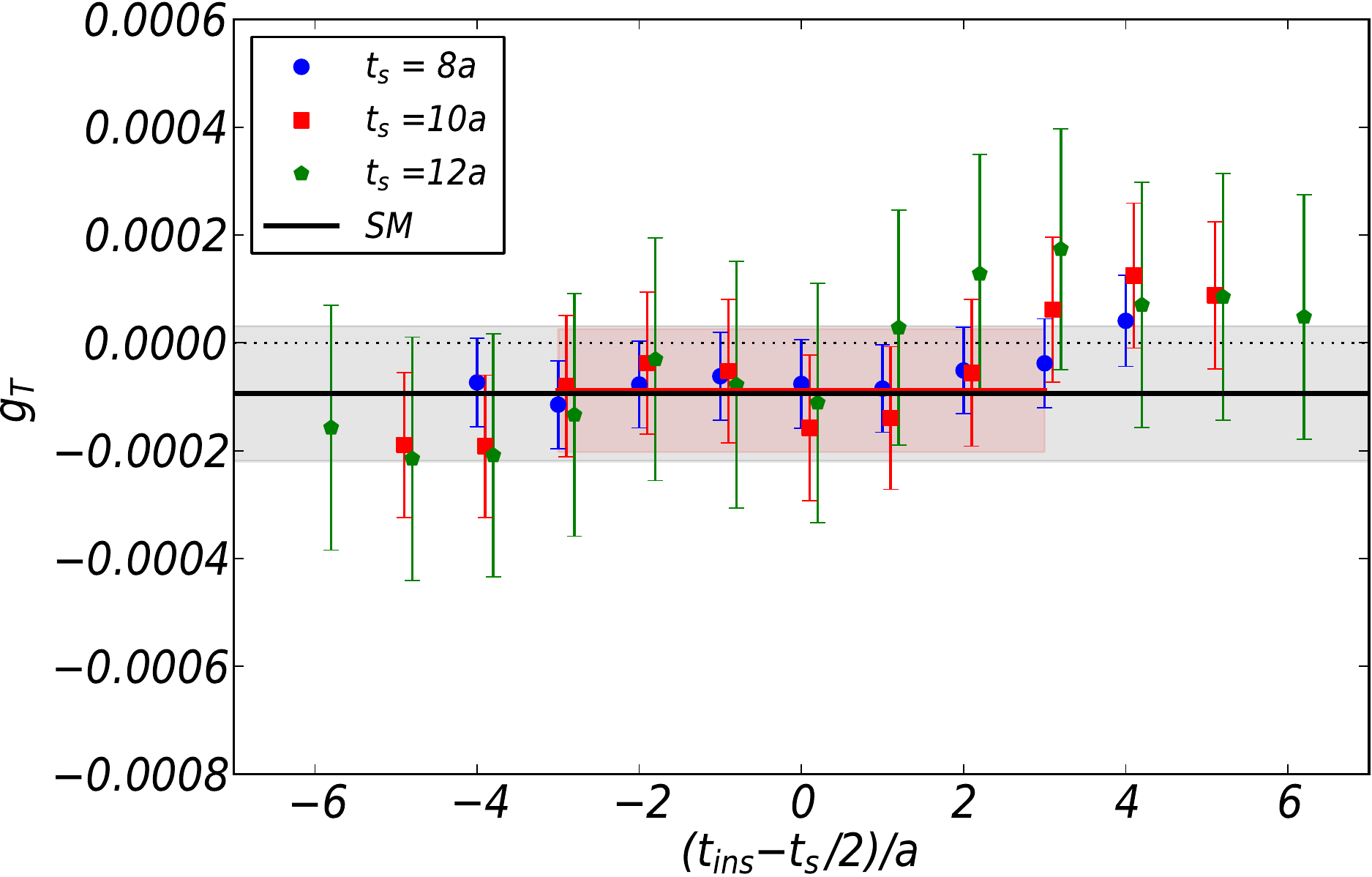} 
      \end{minipage}
      \hspace{0.05\linewidth}
      \begin{minipage}{0.4\linewidth}
        \includegraphics[width=\linewidth,angle=0]{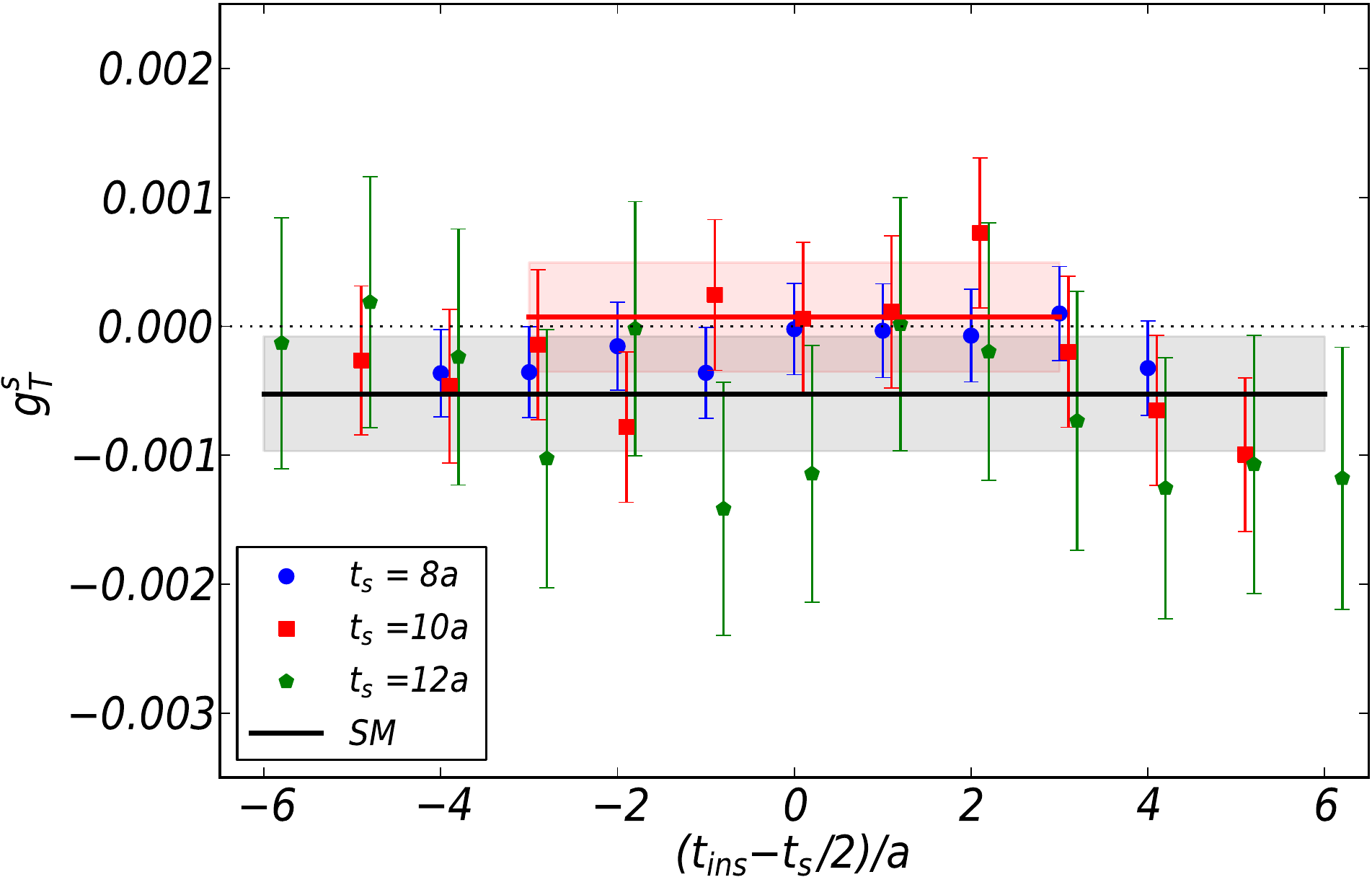} 
      \end{minipage}
      \caption{\footnotesize Plateaux for nucleon disconnected $g_T^{u+d}$ (left) and $g_T^s$ (right). The notation is as in the previous plot, and the results are renormalized without considering the $g_T$ mixing between flavors.
      \label{platgT}}
   \end{center}
   \vspace{-0.5cm}
\end{figure*}

In this first analysis we calculated the isoscalar disconnected and strange contributions to the nucleon $\sigma$-term, $g_A$ and $g_T$, as shown in figs.~\ref{platSig}, \ref{platgA} and \ref{platgT}. Observe
that in the case of the sigma terms, the contributions of the excited states force us to go to large source-sink separations, and the summation method proves critical to find the point were the excited states are
completely suppressed. Other quantities are noisier, but don't seem to be strongly affected by excited states contamination.

\begin{figure*}[h!]
   \begin{center}
      \begin{minipage}{0.35\linewidth}
	    \includegraphics[width=\linewidth,angle=0]{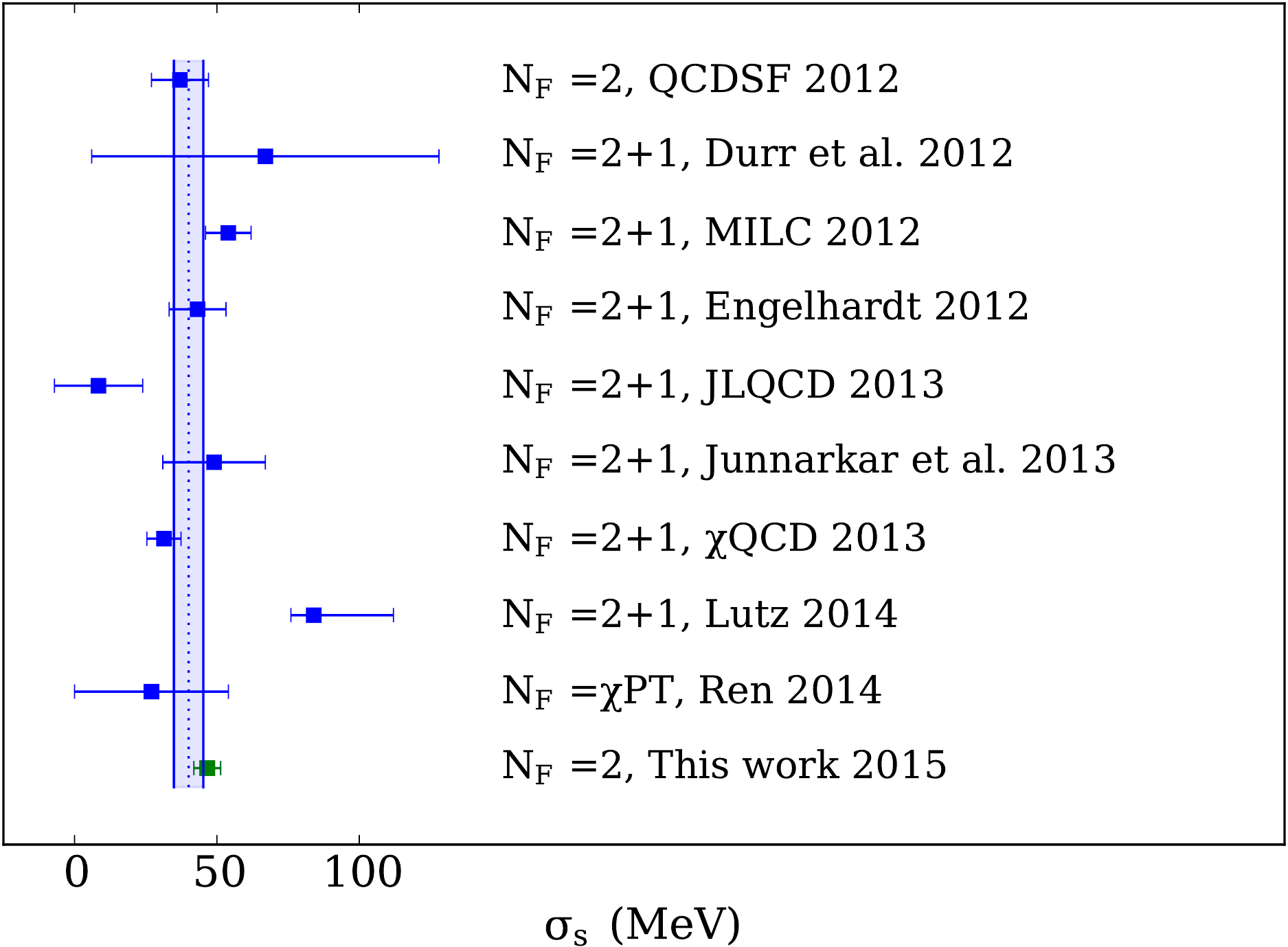}
      \end{minipage}
      \hspace{0.1\linewidth}
      \begin{minipage}{0.35\linewidth}
        \includegraphics[width=\linewidth,angle=0]{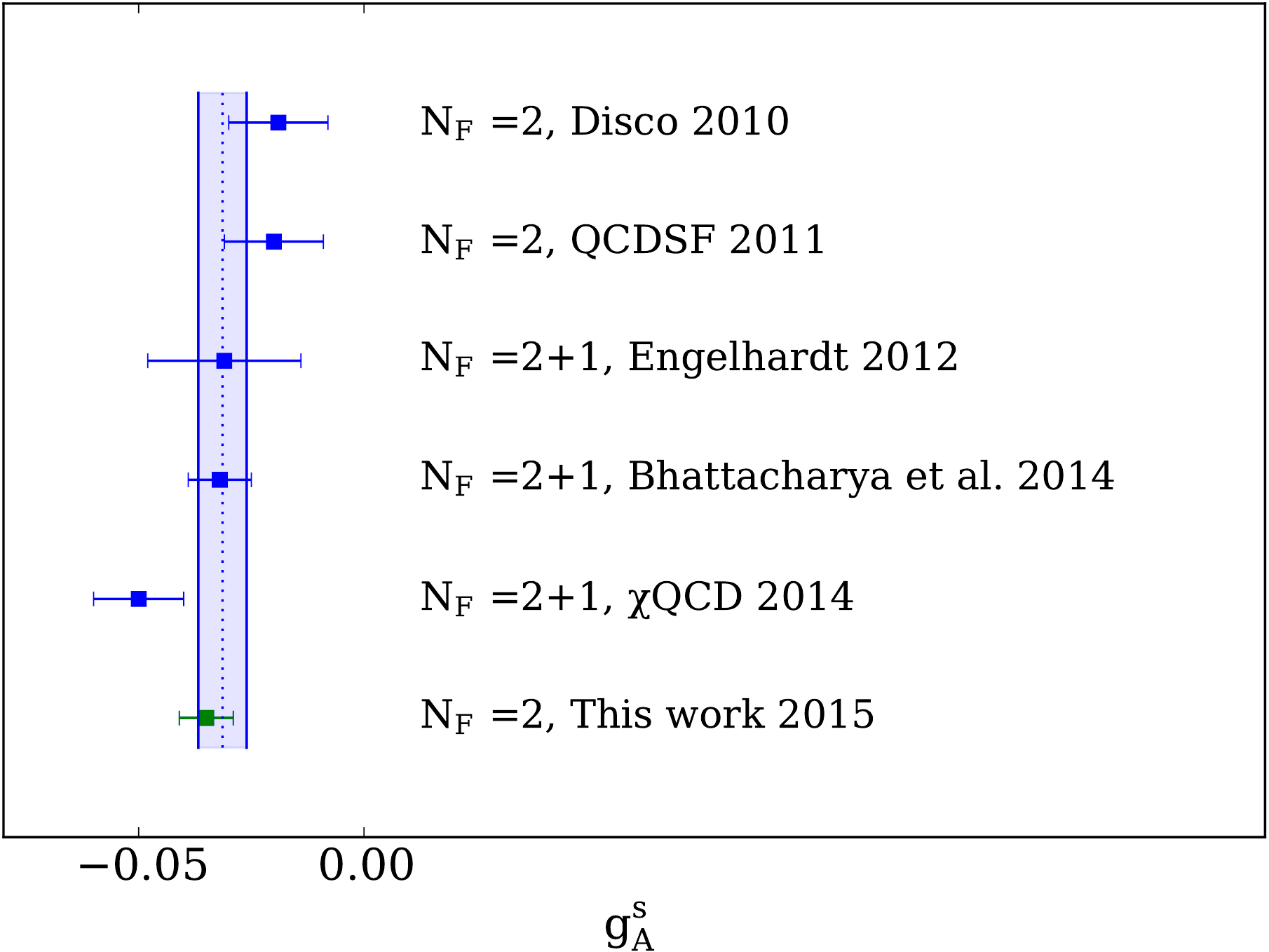} 
      \end{minipage}
      \caption{\footnotesize Comparison of our preliminary calculation with previous results. The central band represents a weighted average that does not include our values. Left: $\sigma_s$. Right: $g_A^s$. \label{compOld}}
   \end{center}
\end{figure*}

For the cases of $\sigma_s$ and $g_A^s$, we also gathered a collection of previous results, extrapolated carefully to the physical point, and compared them to our values. We found a remarkable agreement as summarized in
fig. \ref{compOld}.

\section{Discussion}
\label{sec:con}
In this paper a preliminary computation of disconnected diagrams directly at the physical pion mass is reported. Although finite volume and discretization effects have not been assessed yet, the fact that our results are
in good agreement with previous results that cared about these errors indicates that these effects must be small. We plan in the near future to extend the number of observables analyzed, including one-derivative operators,
as well as to analyze more ensembles to estimate discretization and finite volume effects.

\acknowledgments
This work was supported, in part, by a grant from the Swiss National Supercomputing Centre (CSCS) under project ID s540. Additional computational resources were provided by the Cy-Tera machine at The Cyprus Institute funded by
the Cyprus Research Promotion Foundation (RPF), ${\rm NEAY\Pi O \Delta OMH}$/${\Sigma}$TPATH/0308/31.

\end{document}